\documentclass{article}

\PassOptionsToPackage{numbers,sort&compress}{natbib}
\usepackage{natbib}
\usepackage{graphicx}

\usepackage[utf8]{inputenc} 
\usepackage[T1]{fontenc}    
\usepackage{hyperref}       
\usepackage{url}            
\usepackage{booktabs}       
\usepackage{amsfonts}       
\usepackage{amsmath}
\usepackage{nicefrac}       
\usepackage{microtype}      
\usepackage{xcolor}         
\usepackage{cleveref}

\usepackage[verbose=true,letterpaper]{geometry}

\makeatletter
\newcommand{\@toptitlebar}{
  \hrule height 4\p@
  \vskip 0.25in
  \vskip -\parskip%
}
\newcommand{\@bottomtitlebar}{
  \vskip 0.29in
  \vskip -\parskip
  \hrule height 1\p@
  \vskip 0.09in%
}
\renewcommand{\LARGE}{\@setfontsize\LARGE\@xviipt{20}}
\renewcommand{\@maketitle}{%
  \vbox{%
    \hsize\textwidth
    \linewidth\hsize
    \vskip 0.1in
    \@toptitlebar
    \centering
    {\LARGE\bf \@title\par}
    \@bottomtitlebar
      \def\And{%
        \end{tabular}\hfil\linebreak[0]\hfil%
        \begin{tabular}[t]{c}\bf\rule{\z@}{24\p@}\ignorespaces%
      }
      \def\AND{%
        \end{tabular}\hfil\linebreak[4]\hfil%
        \begin{tabular}[t]{c}\bf\rule{\z@}{24\p@}\ignorespaces%
      }
      \begin{tabular}[t]{c}\bf\rule{\z@}{24\p@}\@author\end{tabular}%
    \vskip 0.3in \@minus 0.1in
  }
}
\makeatother

\title{Machine learning based uncertainty quantification of turbulence model for airfoils}

\author{%
  Minghan Chu\\
  Department of Mechanical and Materials Engineering\\
  Queen's University\\
  \texttt{17MC93@queensu.ca} \\
  \And
  Weicheng Qian\\
  Department of Computer Science\\
  University of Saskatchewan\\
  \texttt{weicheng.qian@usask.ca} \\
}

\begin{document}

\maketitle
\begin{abstract}
Reynolds-averaged Navier-Stokes (RANS)-based transition modeling is widely used in aerospace applications but suffers inaccuracies due to the Boussinesq turbulent viscosity hypothesis. The eigenspace perturbation method can estimate the accuracy of a RANS model by injecting perturbations to its predicted Reynolds stresses. However, there lacks a reliable method for choosing the strength of the injected perturbation, while existing machine learning models are often complex and data craving. We examined two light-weighted machine learning models to help select the strength of the injected perturbation for estimating the RANS uncertainty of flows undergoing the transition to turbulence over a Selig-Donovan 7003 airfoil. On the one hand, we examined polynomial regression to construct a marker function augmented with eigenvalue perturbations to estimate the uncertainty bound for the predicted skin friction coefficient. On the other hand, we trained a convolutional neural network (CNN) to predict high-fidelity turbulence kinetic energy. The trained CNN acts as a marker function that can be integrated into the eigenspace perturbation method to quantify the RANS uncertainty. Our findings suggest that the light-weighted machine learning models are effective in constructing an appropriate marker function that is promising to enrich the existing eigenspace perturbation method to quantify the RANS uncertainty more precisely.
 
\end{abstract}

\section{Introduction}

Turbulence is a phenomenon that manifests over a wide range of length scales. These length scales vary from micro-meters to kilo-meters. The motions at each length scale are coupled and strongly interacting. Thus, any numerical simulation needs to account for these motions. Direct numerical simulation (DNS) computes all scales of motion, large eddy simulation (LES) computes only the large scales while using turbulence models for smaller scales, Reynolds-averaged Navier-Stokes (RANS) uses turbulence models for all scales. Due to this, RANS-based simulations are computationally inexpensive and are widely used in industry and academia for analyzing turbulent flows. However, RANS models are often inaccurate when predicting complex flows due to the turbulent viscosity hypothesis \cite{craft1996development}, especially when modeling the transitional flow regime as widely encountered in turbo-machines.

Less than a decade ago, Iaccarino \emph{et al.} proposed a physics-based eigenspace perturbation approach \cite{emory2013modeling,gorle2019epistemic} to estimate the model-form uncertainty introduced in RANS-based models by quantifying the model-form uncertainty via sequential perturbations to the amplitude (turbulence kinetic energy) \cite{gorle2013framework,cremades2019reynolds}, shape (eigenvalues) \cite{emory2013modeling,gorle2019epistemic}, and orientation (eigenvectors) \cite{iaccarino2017eigenspace,mishra2017uncertainty} of the predicted Reynolds stress tensor. Studies of RANS uncertainty quantification (UQ) are mainly focused on the eigenvalue and eigenvector perturbations, such as RANS uncertainty in flow through scramjets \cite{emory2011characterizing}, aircraft nozzle jets \cite{mishra2017uncertainty}, over steamlined bodies \cite{gorle2019epistemic}, supersonic axisymmetric submerged jet \cite{mishra2017rans}, canonical cases of turbulent flows over a backward-facing step \cite{iaccarino2017eigenspace,cremades2019reynolds}, and benchmark cases of complex turbulent flow \cite{thompson2019eigenvector}. 

When the eigenspace perturbation approach is used to quantify the RANS model uncertainty, it is essential to choose appropriate strength of perturbation. This is often experience-based, which usually introduces the maximum and minimum magnitude of perturbation, resulting in overly conservative perturbations \cite{gorle2012epistemic,iaccarino2017eigenspace,gorle2019epistemic}. Recent machine Learning models can estimate RANS model uncertainty with improved accuracy \cite{xiao2016quantifying,wu2016bayesian,xiao2017random,wang2017physics,wang2017comprehensive,wu2018physics,heyse2021estimating,heyse2021data,zeng2022adaptive}; however, they are often complex and demand a large size of training data. Complex machine learning models not only require additional computational resources in training but also become less comprehensive to researchers. This hinders the understanding and shrinks the room for improvement in the existing theories. Therefore, light-weighted machine learning models are interesting and helpful to incorporate into the physics-based eigenspace perturbation framework to improve the estimates of RANS uncertainty. This study examined two light-weighted machine learning models to bridge the RANS prediction and in-house DNS. Furthermore, we focused on perturbing the turbulence kinetic energy, complementing the eigenvalue and eigenvector perturbations studied by many works.

\section{Methodology}
\label{sec:methodology}

\subsection{Eigenspace Perturbation}
The RANS-based transition model of Langtry and Menter \cite{langtry2009correlation} is used to simulate flows over a SD7003 airfoil at $8^\circ$ angle of attack (AoA). With the Reynolds number based on the cord length of $Re_{c} = 60000$, the flow underwent transition to turbulence on the suction side of the airfoil. Langtry and Menter's model approximates the Reynolds stresses as
\begin{equation}
    \left\langle u_i u_j\right\rangle=\frac{2}{3} k \delta_{i j}-2 v_{\mathrm{t}}\left\langle S_{i j}\right\rangle,
\end{equation}

where $k$ is the turbulence kinetic energy, $\delta_{i j}$ is the Kronecker delta, $\nu_{t}$ is the turbulent viscosity, and $\left\langle S_{i j}\right\rangle$ is the rate of mean strain tensor.
The solution domain is a two-dimensional C-topology grid of $389$ (streamwise) $\times$ $280$ (wall-normal) control volumes. At the inlet, a very low value of turbulence intensity $Tu = 0.03\%$ was assumed. At the outlet, a zero-gradient boundary condition was used for $\left\langle U_i\right\rangle$, $k$, $\omega$, and pressure. At the wall, a no-slip boundary condition was assumed. Grid dependency study showed that finer mesh in the near-wall region did not yield more accurate results. 

The RANS-based transition model is implemented within the OpenFOAM framework and the transient PIMPLE solver was used for pressure-velocity coupling. The solution fields were iterated until convergence: the results of lift and drag coefficients as a function of time were used to track the instant convergence status. At the convergence state, the residuals of energy and momentum dropped more than four orders of magnitude, which required approximately 35000 iterations for all simulations. 

In the eigenspace perturbation approach \cite{emory2013modeling,iaccarino2017eigenspace}, the perturbed Reynolds stresses are defined as

\begin{equation}\label{Eq:Rij_perturb}
        \left\langle u_{i} u_{j}\right\rangle^{*}=2 k^{*}\left(\frac{1}{3} \delta_{i j}+v_{i n}^{*} \hat{b}_{n l}^{*} v_{j l}^{*}\right),
\end{equation}

where $k^{*}$ is the perturbed turbulence kinetic energy, $\hat{b}_{k l}^{*}$ is the perturbed eigenvalue matrix, $v_{i j}^{*}$ is perturbed eigenvector matrix, and $\delta_{i j}$ is the kronecker delta. 

In this study the perturbed turbulence kinetic energy can be defined as

\begin{equation}\label{Eq:Marker_Mk_Method}
    k^{*} = k +\Delta_k = M_{k}, \quad M_{k} \sim f(x,y),
\end{equation}

where $M_{k}$ is a marker function of the $x$ and $y$ coordinate in a computational domain. In this study, the RANS predictions bereft of any perturbations are referred to as ``baseline'' solutions.

\subsection{Projecting a RANS model to a DNS model}

\subsubsection{Polynomial regression}
\begin{figure}[h!]
         \centering
         \includegraphics[width=14cm]{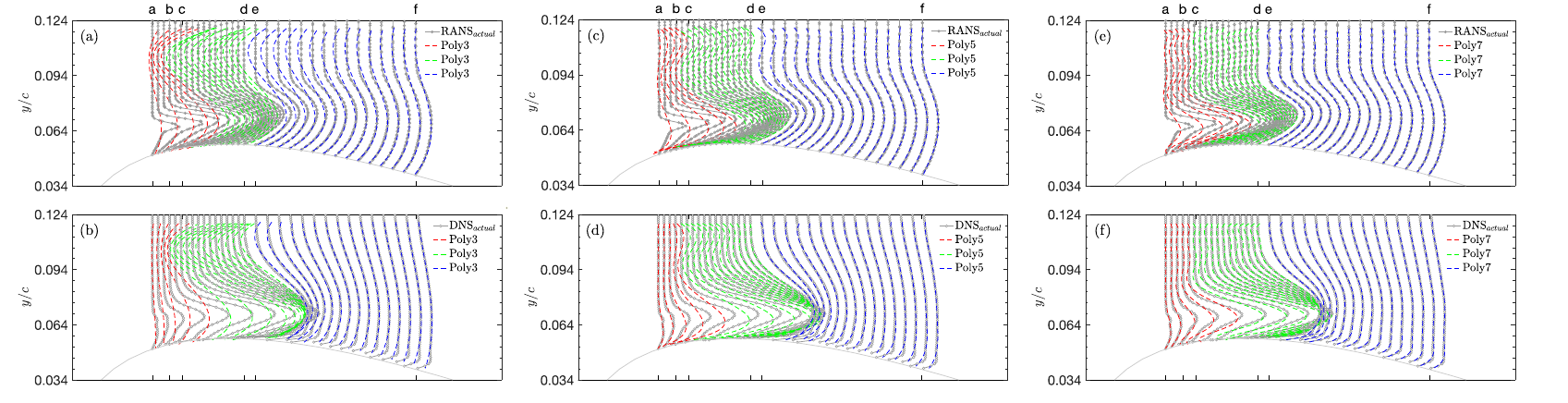}
        \caption{Fitting polynomials of order 3, 5 and 7 to the datasets for RANS and DNS. There are 32 positions on the suction side of the airfoil. $\mathrm{RANS}_{actual}$ represents the baseline prediction. $\mathrm{DNS}_{actual}$ data are included for reference. zone $ab$: $0.14 < x/c < 0.18$; zone $cd$: $0.18 < x/c < 0.3$; zone $ef$: $0.3 < x/c < 0.6$. $c$ is the cord length. }
        \label{fig:Poly_orders.pdf}
\end{figure}

We applied polynomial regressions to fitting the RANS prediction and in-house DNS separately to $n^\mathrm{th}$ order polynomials, with the $n^\mathrm{th}$ order as a hyperparameter, as shown in \ref{fig:Poly_orders.pdf} (a)-(f). From Figs. \ref{fig:Poly_orders.pdf} (a) and (b), the third order polynomials in general are loosely fitted, particularly for the $ab$ and $cd$ zone. As the order of polynomial increases to $5$, it is clear that the $cd$ zone are better fitted. For both the $cd$ and $ef$ zone, the fitted polynomials almost overlap the training datasets. Figs. \ref{fig:Poly_orders.pdf} (e) and (f), most zones are well fitted except for the $ab$ zone shown in Fig. \ref{fig:Poly_orders.pdf} (e). This suggests that the dynamics of turbulence kinetic energy in the $ab$ zone, where a LSB is evolving, are more complex than polynomial.

\subsubsection{Convolutional neural network}
\label{sec:methodology-cnn}
We employed a one-dimensional convolutional neural network (1D-CNN) to learn the projection from RANS estimated function $f^{\scriptstyle\mathrm{RANS}}(x, y)$ to DNS estimated function $f^{\mathrm{DNS}}(x, y)$. For a given $x$, we can rewrite the estimated function as $g_x(k,  y)$. Assuming that there exists a morphism $F$ from $g^{\mathrm{RANS}}_x(k,  y)$ to $g^{\mathrm{DNS}}_x(k,  y)$, then every $x$ and $g_x(k, y)$ is smooth. Our CNN is trained to depict $F$ in the hope of projecting $g^{\mathrm{RANS}}_x(k,  y)$ to $g^{\mathrm{DNS}}_x(k,  y)$. This is conducted by training the paired RANS- and DNS-estimated functions at the selected $x$ coordinates. Taking advantage of the smoothness assumption of $g_x(k, y)$, our 1D-CNN is trained to predict DNS estimated function at $(x, y_\mathrm{target})$ given a series of RANS estimated function at $(x, y_i)$, where $y_i \in [y - \epsilon, y + \epsilon], \epsilon > 0$ belongs to the neighbor of $y_\mathrm{target}$. Our 1D-CNN has four-layers and in total 86 parameters: a single model for all zones at any $x$ to project RANS to DNS. We trained our 1D-CNN with normalized pairs of $\left( g^{\mathrm{RANS}}_x(k,  y), g^{\mathrm{DNS}}_x(k,  y) \right)$ at only three positions $x = 0.4, 0.56, 0.58$ with mean squared error as the loss function and a 80\%--20\% split as training--testing dataset. We validated our trained 1D-CNN by comparing the L1 loss of RANS, denoted as $L^1_c(\texttt{rans}) = \lvert CF^{\mathrm{RANS}}_{k} - CF^{\mathrm{DNS}}_{k} \rvert$ with the L1 loss of 1D-CNN projected RANS, denoted as $L^1_c(\texttt{pred}) = \lvert CF^{\mathrm{CNN}}_{k} - CF^{\mathrm{DNS}}_{k} \rvert$.

\begin{figure}[h!]
         \centering
         \includegraphics[width=8cm]{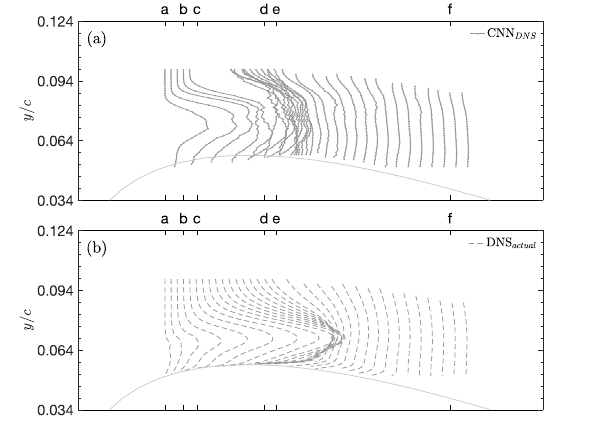}
        \caption{CNN projected DNS (\texttt{CNN\_{DNS}}) compared with ground truth (\texttt{DNS\_{actual}}). There are 32 positions on the suction side of the airfoil.}
        \label{fig:CNN_DNS.pdf}
\end{figure}

\section{Results}
\subsection{Polynomial regression}
\begin{figure}[h!]
         \centering
         \includegraphics[width=10cm]{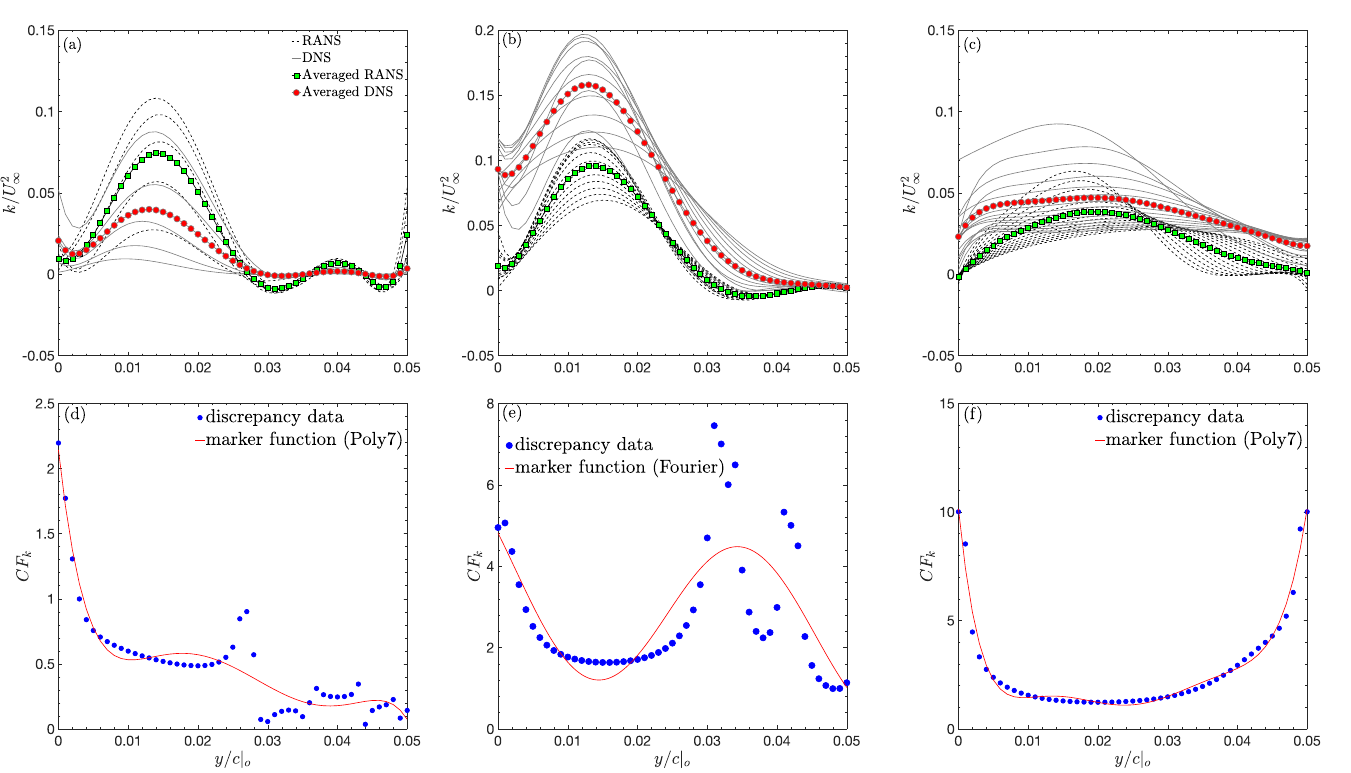}
        \caption{Mean of seventh order polynomials for normalized turbulence kinetic energy (a) - (c), and the corresponding marker function (d) - (f). (a) and (d) zone $ab$; (b) and (e) zone $cd$; (c) and (f) zone $ef$.}
        \label{fig:Discrepancy_Marker.pdf}
\end{figure}

In Figures \ref{fig:Discrepancy_Marker.pdf} (a) - (c), the mean of the seventh-order regression-based normalized turbulence kinetic energy profile as the representative for each zone shows the discrepancy between RANS and in-house DNS. The discrepancy marks the degree of untrustworthiness in the $y /\left.c\right|_o$ direction. Note that all profiles are shifted down to the origin of $y/c$, denoted $y /\left.c\right|_o$. The corresponding marker for each zone is shown in Figs. \ref{fig:Discrepancy_Marker.pdf} (d) - (f). Marker functions can be constructed by fitting appropriate models to the discrepancy data for each zone, i.e., a seventh-order polynomial for the $ab$ and $ef$ zone, and a Fourier series for the $cd$ zone.

\begin{figure}[h!]
         \centering
         \includegraphics[width=6cm]{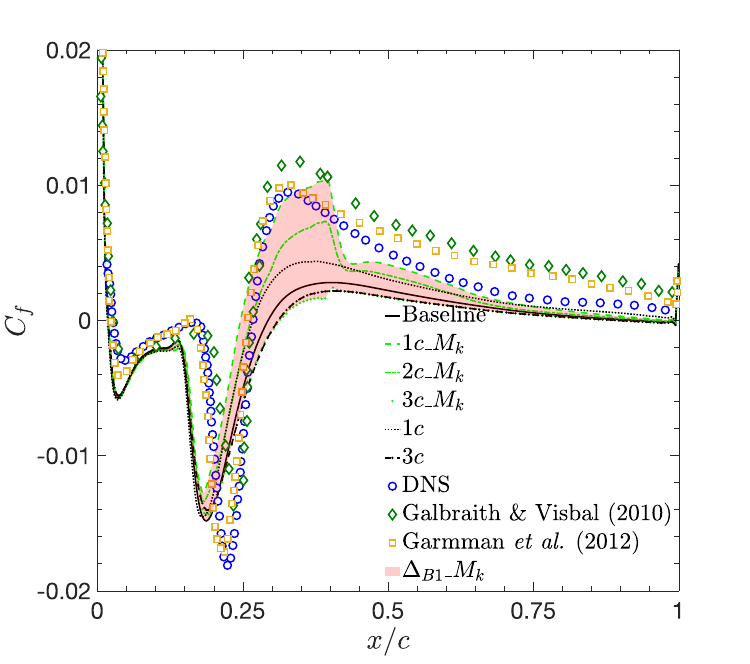}
        \caption{Skin friction coefficient. Displayed are uncertainty bounds for $1c\_M_{k}$, $2c\_M_{k}$ and $3c\_M_{k}$ perturbations (red envelope). $\Delta_{B1}$ stands for $\Delta_{B} = 1$. Profile of the baseline prediction and eigenvalue perturbations ($1c$ and $3c$) are provided for reference.}
        \label{fig:Cf.pdf}
\end{figure}

Augmenting the eigenvalue perturbation \footnote{The strength of eigenvalue perturbation is denoted $\Delta_{B}$, which varies from $0$ to $1$.} with the marker function ($1c\_M_{k}$, $2c\_M_{k}$ and $3c\_M_{k}$) using Eqs. \ref{Eq:Rij_perturb} and \ref{Eq:Marker_Mk_Method}, the estimated model-form uncertainty (red envelope) for the predicted skin friction coefficient is constructed and shown in Fig. \ref{fig:Cf.pdf}. The $1c$ and $3c$ eigenvalue perturbations are included for reference. It is clear that the uncertainty bound successfully encompasses the ILES/LES data of Galbraith \textit{et al.} \cite{galbraith2010implicit} and Garmann \textit{et al.} \cite{garmann2013comparative} for $0.25 < x/c < 0.45$. This region falls into the $cd$ and part of the $ef$ zone, where the LSB is forming and the flow is re-attaching on the wall surface. In comparison to the eigenvalue perturbations, the red envelop exhibits a significant increase in the magnitude of $C_{f}$, showing a tendency to retain the shape of the reference data. This marks a significant improvement in the RANS model prediction for $C_{f}$. The shape of the red envelope is not as smooth as the eigenvalue perturbations, reflecting the effect of spatial variability in $M_{k}$.  

\subsection{CNN}
\begin{figure}[h!]
    \centering
    \includegraphics[width=\textwidth]{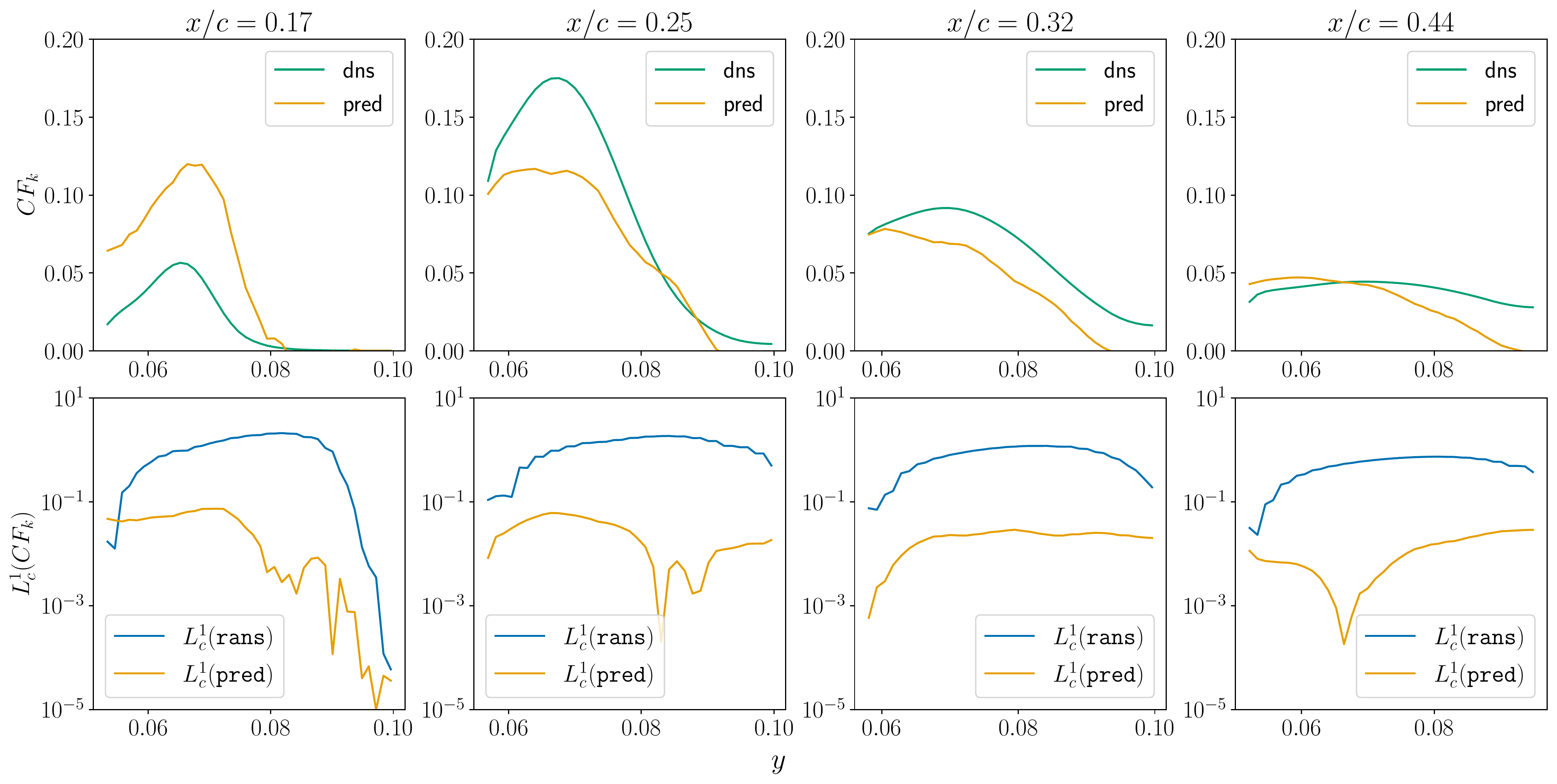}
    \caption{First row: CNN projected DNS (\texttt{pred}) compared with ground truth (\texttt{dns}). Second row: Validation of 1D-CNN by comparing L1 loss between $L^1_c(\texttt{rans})$ and $L^1_c(\texttt{pred})$.}
    \label{fig:cnn-projected-dns-with-rands.pdf}
\end{figure}

The 1D-CNN can predict DNS at any zone given RANS, thus acting as the marker function $M_{k}$ in Eq. \ref{Eq:Marker_Mk_Method}. From Figs. \ref{fig:CNN_DNS.pdf} (a) and (b), the CNN predicted DNS profile for $k$ shows overall good resemblance to the DNS dataset, although an over-prediction exists at the beginning of the $ab$ zone. 
From the Fig. \ref{fig:cnn-projected-dns-with-rands.pdf}, the series of CNN predicted DNS profiles in the first row are then smoothed with the moving average with a window size of six consecutive estimations. Our CNN predicted DNS profiles resemble the ground truth DNS despite being trained with only a few pairs of RANS and DNS results. From Fig. \ref{fig:cnn-projected-dns-with-rands.pdf}, the discrepancy in general reduces as the flow proceeds further downstream. Consequently, the CNN predicted DNS given the RANS estimated function acts as the marker function $M_k$ in Eq. \ref{Eq:Marker_Mk_Method}. From the Fig. \ref{fig:cnn-projected-dns-with-rands.pdf}, the second row shows the computed error of the baseline solution and the CNN predicted DNS, and it is clear that the error for CNN predicted DNS is significantly reduced in magnitude compared to that for the baseline solution. 

\section{Conclusion}
\label{sec:Conclusion}

In this study, the flow being considered is over a SD7003 airfoil at $8^\circ$ angle of attack and the Reynolds number based on the cord length of $Re_{c} = 60000$. A laminar separation bubble evolves within the separated boundary layer whereby the flow undergoes transition to turbulence. The goal of this study was to assess the effect of the polynomial regression and the CNN approaches on the RANS model uncertainty quantification. The learning algorithms need to be coupled to the eigenspace perturbation approach of Emory \textit{et al.} \cite{emory2013modeling} that is implemented within the OpenFOAM framework to construct a marker function for the turbulence kinetic energy perturbation. 

Seventh-order polynomial regressions overall captures the discrepancy in the predicted turbulence kinetic energy between RANS and in-house DNS. Correspondingly, the the marker function is augmented with the  eigenvalue perturbation to significantly increase the uncertainty bound for $C_{f}$. Around the peak of the $C_{f}$ curve, the uncertainty bound successfully encompassed the reference data in the aft portion of the LSB and the re-attachment point.  

To the best of our knowledge, we are among the first to examine the projection from RANS to DNS using the CNN approach. Our experiment results suggest that the CNN approach can help us project the RANS estimated marker function to the in-house DNS data. A projection that can approximate the in-house DNS reasonably well from RANS might exist independent of $x$. Our methodology can be easily extended to analyze flows over different types of airfoils.

Our findings are subject to following limitations. The results of our machine learning based uncertainty quantification are subject to the RANS-based transition model of Langtry and Menter when simulating flows over the SD7003 airfoil at $8^\circ$ angle of attack (AoA). With the Reynolds number based on the cord length of $Re_{c} = 60000$. Future work may include evaluating more machine learning models in generating marker functions with different types of airfoils, as well as integrating the CNN approach into the eigenspace perturbation framework to result in accurate perturbations and hence accurate RANS UQ.

\section{Impact statement}
Transitional flow is important because the transitional flow regime is frequently encountered in aerospace applications, while this flow regime is hard to model. LES and DNS can provide high fidelity solution, but the calculations are often expensive. Therefore, the estimation of the RANS uncertainty is a valuable alternative tool for improving the RANS predictive capability in engineering applications. More expensive LES or DNS would only be adopted if the RANS uncertainty is too large. A recent physics-based eigenspace peturbation method evaluates the accuracy of RANS model for practical usage based on experts selected perturbations, which is less reliable in terms of giving inaccurate strength of perturbation. We proposed a machine-learning based eigenspace perturbation method that can effectively increase the precision of the estimates of RANS uncertainty and build confidence in aeronautical simulations by which risky decision-making is always accompanied. Our method is highly generalizable to a variety of flow scenarios. In addition, the method can also be employed to shed lights on predicting deviations of RANS from DNS, promising a correction to RANS to improve its accuracy.


\medskip

{
\small

}

\bibliography{bib}
\appendix

\end{document}